\documentclass[aps,prb,reprint,unsortedaddress,superscriptaddress]{revtex4-1}

\usepackage{amsmath,amssymb}

\usepackage{mediabb}
\usepackage{graphicx}

\usepackage{bm}
\usepackage{color}
\begin{document}

\title{Spin Berry phase in anisotropic topological insulators}

\begin{abstract}
Three-dimensional topological insulators are characterized by the presence of
protected gapless spin helical surface states.
In realistic samples these surface states are extended 
from one surface to another, covering the entire sample.
Generally, 
on a curved surface of a topological insulator 
an electron in a surface state acquires a spin Berry phase 
as an expression of the constraint
that the effective surface spin must follow
the tangential surface of real space geometry.
Such a Berry phase adds up to $\pi$ when
the electron encircles, e.g., once around a cylinder.
Realistic topological insulators compounds are also often
layered, i.e., are anisotropic.
We demonstrate explicitly the existence of such a $\pi$ Berry phase
in the presence and absence (due to crystal anisotropy) of cylindrical symmetry, 
that is, regardless of fulfilling the spin-to-surface locking condition. 
The robustness of
the spin Berry phase $\pi$ against cylindrical symmetry breaking is
confirmed numerically using a tight-binding model implementation of
a topological insulator nanowire penetrated by a $\pi$-flux tube.
\end{abstract}

\date{\today}

\author{Ken-Ichiro Imura}
\affiliation{Department of Quantum Matter, AdSM, Hiroshima University, Higashi-Hiroshima 739-8530, Japan}
\author{Yositake Takane}
\affiliation{Department of Quantum Matter, AdSM, Hiroshima University, Higashi-Hiroshima 739-8530, Japan}
\author{Akihiro Tanaka}
\affiliation{National Institute for Materials Science, Tsukuba 305-0047, Japan}

\maketitle

\section{Introduction}
Consider the situation where 
the lower half of a three dimensional space is occupied by
a topological insulator with 
$\mathbb Z_2$ index $\nu=1$ and the rest is a vacuum ($\nu=0$). 
Then as the most characteristic feature of the topological insulating state, 
a metallic surface state appears 
at its interface with the vacuum. 
\cite{Moore_review, Hasan_Kane, Fu_Kane}
An interesting variant of this scenario explored by a recent Aharonov-Bohm measurement 
on Bi$_2$Se$_3$ nanowire 
\cite{AB_exp}
and subsequent theoretical analyses
\cite{Vishwanath_PRB, Vishwanath_PRL, Mirlin, Moore, Franz, ITT}
is that the surface metallic state is not only protected by time-reversal symmetry,
but it shows another characteristic feature when the surface is deformed,
say, into a cylinder (see FIG. 1)
--- the manifestation of the spin Berry phase. 
Depending on how the surface is deformed into a cylinder, i.e.,
whether the topological material fills either the inside or the outside of the cylinder, 
the system can be regarded as either a nanowire or a linear aperture penetrating 
an otherwise surfaceless topological insulator. 
In a recent work 
\cite{ITT}
we have chosen the latter as a starting point for studying the nature of
topologically protected helical modes along a dislocation line. 
\cite{Ran, Teo_Kane}
Effects of a finite size (a finite radius of the cylinder)
combined with the presence of a nontrivial spin Berry phase was shown to play an essential role 
in protecting the 1D helical modes. 
\cite{Vishwanath_PRB, Vishwanath_PRL, Mirlin, ITT}

\begin{figure}
\begin{center}
\includegraphics[width=8cm]{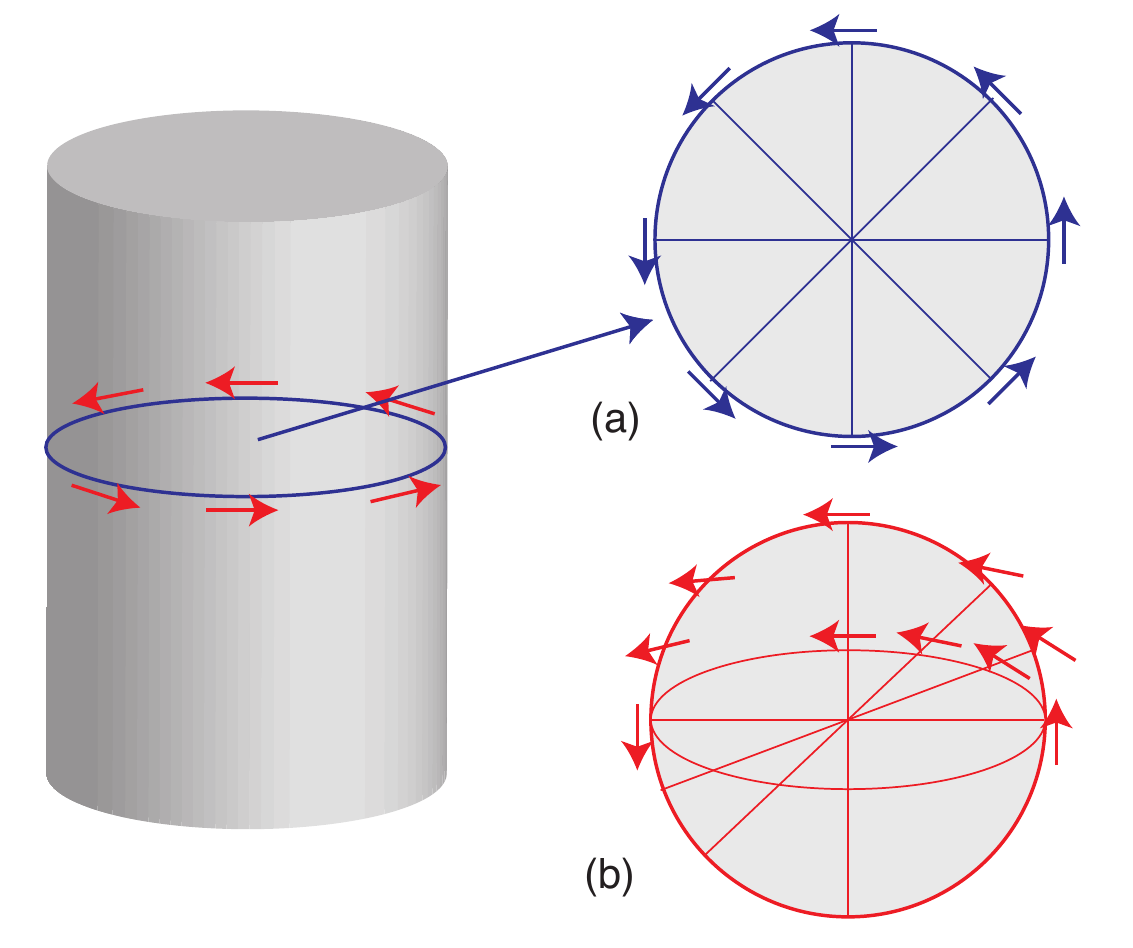}
\end{center}
\caption{Cylindrical surface of a topological insulator nanowire
and a typical spin configuration of a surface electronic state (left).
Top view (a cross section is shown) of the spin configuration (right)
in the (a) presence and (b) absence of cylindrical symmetry.
In case (b) the spin follows the tangential plane of an
auxiliary elliptic surface.}
\label{spec_screw}
\end{figure}

The appearance of spin Berry phase $\pi$ is a characteristic feature of topological insulator 
surface state, 
\cite{berry_pi}
distinguishing it from, e.g., a carbon nanotube, 
\cite{R}
another 2D gapless Dirac system (i.e., graphene) rolled up into a cylinder. 
These two Dirac systems both involve
an effective spin degree of freedom
appearing in the low-energy effective Hamiltonian.
The physical origins of these spin degrees of freedom 
are different in the two cases;
in the case of graphene (or carbon nanotube)
it is the sub-lattice structure of hexagonal lattice,
whereas in the present example 
it is essentially a genuine electron spin. 
These two effective spin degrees of freedom,
despite their very different nature, 
play a similar role in determining the transport characteristics
of the surface states on a {\it flat} surface.
This is, however, no longer the case
when the surface is {\it curved}.
The sub-lattice pseudo-spin of graphene is
insensitive to warping of the 2D plane.
The effective spin on the surface of a topological insulator is, on the contrary, 
constrained to lie in-plane to the surface.
\cite{spin_locking}
This constraint is the origin of the spin Berry phase $\pi$
characteristic to the topological insulator surface state.
\cite{Vishwanath_PRB, Vishwanath_PRL, Mirlin}
One of the purposes of the present work is to demonstrate through explicit examples how the information encoded within the bulk Hamiltonian manifests itself in the surface effective Hamiltonian in the form of a nontrivial spin Berry phase.
In the course of deriving the spin Berry phase,
we also establish an unambiguous correspondence between
the effective spin degree of freedom appearing in the 2D surface Dirac Hamiltonian
and the original real spin embedded into the 3D bulk effective Hamiltonian.

A second motivation of this work is to explore the consequences of the anisotropy of 
topological insulators on the surface states, especially on the spin Berry phase.
A rectangular nanowire made of such asymmetric compounds have
surfaces of different symmetries.
For example, 
in the Bi$_2$Se$_3$ nanowire studied in Ref. \cite{AB_exp}
the surfaces orthogonal to the $c$-axis is rotationally symmetric,
whereas surfaces parallel to the $c$-axis is not.
Correspondingly, the surface Dirac cones are symmetric in the former,
whereas distorted in the latter.
The slope of the energy dispersion (Fermi velocity) also differs.
What happens to the electron when it goes through (or gets reflected? by)
junctions between two surfaces of different character?
What is the fate of the Berry phase $\pi$ when an electron goes around the nanowire
passing by several of such junctions?
Such effects of anisotropy and multiple surface geometry
will be also important in a similar Josephson-Majorana geometry
involving metallic surface and Majorana bound states.
\cite{Ioselevich}

This paper is organized as follows:
we first demonstrate (Sec. II) by a simple analytic calculation that 
in the presence of anisotropy the surface effective spin 
is no longer strictly locked in-plane to the tangential plane on a curved surface,
but it has generally a finite out-of-plane component.\cite{Thierry}
We also show that only the global Berry phase $\pi$ which an electron acquires
when it winds around a cylinder
is robust against such asymmetry of the crystal.
This point is further confirmed numerically 
by implementing a topological insulator nanowire 
penetrated by a $\pi$-flux tube
as a tight-binding model on a square lattice (Sec. III).

\section{Derivation of the spin Berry phase}
Let us first derive the spin Berry phase $\pi$ 
{\it directly} from a bulk 3D effective Hamiltonian.
In contrast to Refs. \cite{Vishwanath_PRB, Vishwanath_PRL,Mirlin},
here, we choose to go back to the 3D bulk effective Hamiltonian,
and derive the spin Berry phase directly from the 3D Hamiltonian.

\subsection{Cylindrical nanowire in parallel with the crystal growth axis}
Let us start with the case in which the system has a cylindrical symmetry,
i.e., a bulk insulating state is confined inside a cylinder of a radius $R$: 
$\sqrt{x^2+y^2} \le R$,
directed along a crystal growth axis ($c$-axis)
perpendicular to the stacking layers.
To describe the bulk insulating state, 
which is in contact with the vacuum outside the cylinder,
we take the following effective Hamiltonian,\cite{Liu_nphys, Liu_PRB, Shen_NJP}
\begin{equation}
\label{ham_bulk}
H_{3D} =
\left[
\begin{array}{cccc}
M & B k_z & 0 & A k_- \\
B k_z & -M  & A k_- & 0 \\
0 & A k_+ &  M & - B k_z \\
A k_+ & 0 & - B k_z & -M
\end{array}
\right].
\end{equation}
This Hamiltonian describes a 3D Dirac system with a mass
parameter $M=M_0+M_2 (k_x^2 + k_y^2 + k_z^2)$.
The nature of the insulating state, i.e., whether it is $\mathbb Z_2$ trivial or not,
is determined by the sign of $M_0/M_2$.
When $M_0/M_2 <0$, the insulating state is $\mathbb Z_2$ nontrivial: $\nu=1$
(the gap is inverted) exhibiting a single gapless surface Dirac cone, whereas
when $M_0/M_2 <0$, $\mathbb Z_2$ index is $\nu=0$ and the gap is normal.
We regard here
$A$ and $B$ to be constant;
$A=A_0$, $B=B_0$ in the parametrization of Ref. \cite{Liu_PRB}.
Note that the same is assumed in the derivation of flat surface states
in Ref. \cite{Liu_PRB}.
The anisotropy of the crystal is reflected in the asymmetry between 
$A$ and $B$;
according to Table IV of Ref. \cite{Liu_PRB},
$B_0$ is generally smaller then $A_0$,
and in the case of Bi$_2$Te$_3$ smaller by one order of magnitude.
The structure of the Hamiltonian (\ref{ham_bulk}) may become clearer
in the follwoing symbolic form:
\begin{equation}
\label{ham_bulk_symb}
H = M \tau_z + B k_z  \tau_x \sigma_z + A \tau_x (k_x \sigma_x + k_y \sigma_y),
\end{equation}
where
$\bm \tau = (\tau_x, \tau_y, \tau_z)$ and
$\bm \sigma = (\sigma_x, \sigma_y, \sigma_z)$ are two sets of Pauli matrices
representing, respectively, an orbital and a spin degrees of freedom.

To identify the gapless surface states in the cylindrical geometry,
we first decompose, 
in the spirit of Ref. \cite{Liu_PRB, Shen_NJP},
the bulk 3D effective Hamiltonian (\ref{ham_bulk})
into two parts:
\begin{equation}
\label{decomp}
H = H_\perp (k_r) + H_\parallel (k_\phi, k_z),
\end{equation}
where $H_\perp = H|_{k_\phi=k_z=0}$.
$k_r$ and $k_\phi$ are components of the crystal momentum
conjugate to the cylindrical coordinates:
\begin{equation}
\label{theta_def}
r=\sqrt{x^2+y^2},\ \ \
\phi=\arctan {y\over x}.
\end{equation}
$H_\perp$ is a Hamiltonian at the $\Gamma$-point, and reads explicitly,
\begin{eqnarray}
\label{ham_perp_symb}
H_\perp &=& M_\perp \tau_z + A k_r \tau_x (\sigma_x \cos \phi + \sigma_y \sin \phi)
\\
&=&
\left[
\begin{array}{cccc}
M_\perp & 0 & 0 & A e^{-i\phi} k_r \\
0 & -M_\perp & A e^{-i\phi} k_r & 0 \\
0 & A e^{i\phi} k_r & M_\perp & 0 \\
A e^{i\phi} k_r & 0 & 0 & -M_\perp
\end{array}
\right].
\label{ham_perp}
\end{eqnarray}
For a later use, note also that the remaining $H_\parallel$ becomes
\begin{equation}
\label{ham_para}
H_\parallel =
\left[
\begin{array}{cccc}
M_\parallel & B k_z & 0 & -i A e^{-i\phi} k_\phi \\
B k_z & -M_\parallel  & -i A e^{-i\phi} k_\phi & 0 \\
0 & i A e^{i\phi} k_\phi & M_\parallel & - B k_z \\
i A e^{i\phi} k_\phi & 0 & - B k_z & -M_\parallel
\end{array}
\right].
\end{equation}
In Eqs. (\ref{ham_perp}) and (\ref{ham_para})
we have decomposed the mass term into
$M_\perp \simeq M_0+M_2 k_r^2$ and
$M_\parallel = M_2 (k_\phi^2 + k_z^2)$.
Of course, the Laplacian in the cylindrical coordinates
has another contribution, $(1/r)\partial / \partial r$.
Here, we neglect this first-order derivative term, keeping the term
$\partial^2 / \partial r^2$,
which is {\it a posteori} justified,
since the penetration depth $\lambda$ of the surface state is much smaller than
the radius of the cylinder; 
$R\gg \lambda$.

We then consider a solution of the eigenvalue equation,
\begin{equation}
H_\perp |\psi \rangle = E_\perp |\psi \rangle
\end{equation}
of the form,
\cite{Shen_PRL, Imura_PRB, Liu_PRB, Shen_NJP}
\begin{equation}
|\psi \rangle \sim e^{\lambda (r-R)},
\label{lambda}
\end{equation}
i.e., 
$k_r = -i\lambda$ (we keep only $\lambda >0$).
For a given $E_\perp$, one finds four independent solutions of this form,
$|\psi \rangle = |\psi_j \rangle$ ($j=1,2,3,4$).
Then one composes a linear combination of these four solutions,
\begin{equation}
\label{sol_gen_hypo}
|\psi \rangle = \sum_{j=1}^4 c_j |\psi_j \rangle,
\end{equation}
for satisfying the boundary condition:
\begin{equation}
\label{bc}
| \psi \rangle_{r=R} =
\left[
\begin{array}{c}
0 \\ 0\\ 0\\ 0  
\end{array}
\right].
\end{equation}
The boundary condition (\ref{bc}) gives a restriction to the spinor
part of the wave function $|\psi \rangle$, allowing for
explicitly writing down its two independent bases.

To find the eigenspinors, which compose the spinor part of $|\psi_j\rangle$,
one needs to diagonalize $H_\perp$.
In view of its specific form
written symbolically as in Eq. (\ref{ham_perp_symb}),
one may diagonalize its real spin part first.
Namely, one can partially diagonalize $H_\perp$ as,
\begin{equation}
\label{ham_perp_pm}
H_\perp |\bm r  \pm \rangle = 
\left( M_\perp \tau_z \pm A k_r \tau_x \right) |\bm r \pm \rangle,
\end{equation}
in terms of the following (real) spin eigenstates pointed in the
radial direction $\hat{\bm r}=(\cos \phi, \sin \phi)$;
\begin{equation}
\label{r_pm}
|\bm r + \rangle = {1\over \sqrt{2}}
\left[
\begin{array}{c}
1 \\ e^{i\phi} 
\end{array}
\right],\ \ 
|\bm r - \rangle =
 {1\over \sqrt{2}}
\left[
\begin{array}{c}
1 \\ - e^{i\phi} 
\end{array}
\right].
\end{equation}
Here, we have {\it chosen} these eigenspinors single-valued.
It is possible, of course, to take them double valued,
but the two choices turn out to be completely equivalent.
The advantage of the single-valued choice is that
the Berry phase becomes explicit in the surface effective Hamiltonian; see Eq. (\ref{ham_surf}).
Whether one chooses one set of $|\bm r \pm \rangle$ or the other,
they compose a set of bases diagonalizing real spin states
pointed in the direction of $\hat{\bm r}$.

The remaining orbital ($\tau$-) part can be also diagonalized as,
\begin{eqnarray}
H_\perp 
|\lambda \pm \rangle
|\bm r  \pm \rangle
&=&
E_\perp
|\lambda \pm \rangle
|\bm r \pm \rangle,
\nonumber \\
|\lambda \pm \rangle &=&
\left[
\begin{array}{c}
1 \\ 
\pm i (E_\perp -M_\perp) / (\lambda A)
\end{array}
\right].
\label{ham_perp_sol}
\end{eqnarray}
In the second line we took $k_r = -i\lambda$
explicitly into account.
For a given energy $E_\perp$
satisfying
$E_\perp^2 = M_0 - M_2 \lambda^2$,
there are four possible solutions for the penetration depth, 
$\lambda= \pm \lambda^{(\pm)}$,
of which we keep only the two positive solutions,
$\lambda= \lambda^{(\pm)}$.
For each of $\lambda= \lambda^{(\pm)}$ 
we have two independent base spinors $|\lambda \pm \rangle$;
we have in total four independent solutions,
constituting the general solution (\ref{sol_gen_hypo}). To be explicit,
the general solution (\ref{sol_gen_hypo}) reads explicitly,
\begin{eqnarray}
\label{sol_gen}
|\psi \rangle =
\left[
c_1 |\lambda^{(+)} +\rangle |\bm r +\rangle +
c_2 |\lambda^{(+)} -\rangle |\bm r -\rangle
\right]
e^{\lambda^{(+)} (r-R)}
\nonumber \\
+\left[
c_3 |\lambda^{(-)} +\rangle |\bm r +\rangle +
c_4 |\lambda^{(-)} -\rangle |\bm r -\rangle
\right]
e^{\lambda^{(-)} (r-R)}.
\end{eqnarray}
Notice that in Eq. (\ref{ham_perp_sol}) $E_\perp$ and
 $M_\perp$ are also functions of $\lambda$;
$E_\perp = E_\perp (\lambda^{(\pm)})\equiv E_\perp^{(\pm)}$,
$M_\perp = M_\perp (\lambda^{(\pm)}) \equiv M_\perp^{(\pm)}$.
The last step is to impose the boundary condition (\ref{bc})
to Eq. (\ref{sol_gen}).

Since the two base spinors $|\bm r +\rangle$ and $|\bm r +\rangle$
subtend orthogonal real spin subspaces,
the boundary condition (\ref{bc})
requires that
\begin{eqnarray}
c_1 |\lambda^{(+)} +\rangle + c_3 |\lambda^{(-)} +\rangle &=& 
\left[
\begin{array}{c}
0 \\ 0  
\end{array}
\right],
\nonumber \\
c_2 |\lambda^{(+)} -\rangle + c_4 |\lambda^{(-)} -\rangle &=& 
\left[
\begin{array}{c}
0 \\ 0  
\end{array}
\right],
\label{E=0}
\end{eqnarray}
independently hold. This means that the spinor part of Eq. (\ref{sol_gen}) 
can be expressed solely in terms of the eigen spinors, say, 
with $\lambda = \lambda^{(+)}$, i.e., as a linear combination of
$|\lambda^{(+)} +\rangle$ and $|\lambda^{(+)} -\rangle$.
The first line of Eq. (\ref{E=0}) implies,
\begin{equation}
\det
\left[
\begin{array}{cc}
- i\lambda^{(+)} A & - i\lambda^{(-)} A \\ 
E_\perp^{(+)} -M_\perp^{(+)} & E_\perp^{(-)} - M_\perp^{(-)}
\end{array}
\right]=0.
\end{equation}
One can verify that this holds true only when the two conditions:
(i) $M_0 M_2 < 0$ (the system is in the $\nu=1$ phase) and 
(ii) $E_\perp =0$ are simultaneously satisfied.
Substitute $E_\perp =0$ into Eq. (\ref{ham_perp_sol}),
and notice that $M_\perp= - \lambda A$ is 
the only choice consistent with the requirement that
$\lambda>0$, if one defines the parameters such that $A>0$ and $M_2>0$.

Taking all these into account
one can express the solution of the boundary problem as,
\begin{eqnarray}
|\psi \rangle &=&
\left[
c_1 |\lambda^{(+)} +\rangle |\bm r +\rangle +
c_2 |\lambda^{(+)} -\rangle |\bm r -\rangle
\right]
\rho (r)
\label{sol_bc}
\\
&=&
\left[
{c_1 \over 2}
\left[
\begin{array}{c}
1 \\ i \\ e^{i\phi} \\ i e^{i\phi}
\end{array}
\right]
+
{c_2 \over 2}
\left[
\begin{array}{c}
1 \\ -i \\ - e^{i\phi} \\ i e^{i\phi}
\end{array}
\right]
\right]
\rho (r)
\nonumber \\
&\equiv&
c_1 |\bm r +\rangle\rangle + c_2 |\bm r - \rangle\rangle
\label{sol_bc_1}
\end{eqnarray}
where 
\begin{equation}
\rho (r) \simeq
{\sqrt{\lambda_+\lambda_-(\lambda_+ +\lambda_-) / \pi R}
\over |\lambda_+ - \lambda_-|}
\left[
e^{\lambda_+ (r-R)} - e^{\lambda_- (r-R)}
\right],
\label{rho_r}
\end{equation}
($\lambda_\pm \ll R$ assumed) and,
\begin{equation}
\label{lambda_pm}
\lambda_\pm ={A \pm \sqrt{A^2 + 4 M_0 M_2} \over 4 M_2}.
\end{equation}

The four-component eigenspinors $|\bm r\pm \rangle\rangle$
introduced in Eq. (\ref{sol_bc_1}) describe
electronic states localized in the vicinity of the surface.
The effective 2D surface Hamiltonian $H_{2D}$ is obtained by
calculating the matrix elements of $H_\parallel$ in terms of these
$| \bm r\pm \rangle\rangle$, i.e.,
\begin{equation}
(H_{2D})_{\pm\pm}=
\langle\langle \bm r \pm| H_\parallel | \bm r\pm \rangle\rangle.
\end{equation}
By an explicit calculation, $H_{2D}$ is found to be
\begin{eqnarray}
\label{ham_surf}
&&H_{2D} = 
\\
&&\left[
\begin{array}{cc}
0 & -i B k_z + {A\over R} \left(-i{\partial \over \partial\phi}+{1\over 2}\right) 
\\ 
i B k_z + {A\over R} \left(-i{\partial \over \partial\phi}+{1\over 2}\right) & 0
\end{array}
\right],
\nonumber
\end{eqnarray}
where we have used,
$k_\phi \simeq -i (1/R) \partial / \partial\phi$,
since $R \gg \lambda$, i.e., only $r \simeq R$ is relevant.
Two factors $1/2$ which have appeared in the off-diagonal elements
of Eq.  (\ref{ham_surf}) are the spin Berry phase terms, 
which lead to a $\pi$-phase shift when an electron 
goes around the cylinder.
This Berry phase term can be eliminated from the eigenvalue
equation for $H_{2D}$
\begin{equation}
\label{surf_eigen}
H_{2D} 
\left[
\begin{array}{c}
c_1 \\ c_2
\end{array}
\right] 
= E_\parallel
\left[
\begin{array}{c}
c_1 \\ c_2
\end{array}
\right] 
\end{equation}
by introducing a singular gauge transformation,
\begin{equation}
\label{SGT}
\bm c=
\left[
\begin{array}{c}
c_1 \\ c_2
\end{array}
\right] 
= e^{-i\phi /2} 
\left[
\begin{array}{c}
\chi_1 \\ \chi_2
\end{array}
\right].
\end{equation}
In the transformed $\chi$-basis,
the surface Hamiltonian takes a simple Dirac form without the Berry phase,
\begin{equation}
\label{ham_surf_w/o}
H_{2D}^{(\bm \chi)}
= A \sigma_x k_\phi + B \sigma_y k_z,
\end{equation}
whereas the corresponding eigenspinors,
\begin{equation}
\label{chi}
\bm \chi =
\left[
\begin{array}{c}
\chi_1 \\ \chi_2
\end{array}
\right],
\end{equation}
become double-valued.

It is suggestive to express $\bm \chi$
explicitly in terms of a set of
polar coordinates defined in the effective surface
spin space.
By introducnig the parameters as,
\begin{equation}
\cos \eta = {A k_\phi \over \sqrt{A^2 k_\phi^2 + B k_z^2}},\ \
\sin \eta =  {B k_z \over \sqrt{A^2 k_\phi^2 + B k_z^2}},
\end{equation}
the eigenstates of Eq. (\ref{ham_surf_w/o}),
corresponding to the eigenenergies,
\begin{equation}
E_\parallel = \pm \sqrt{A^2 k_\phi^2 + B k_z^2},
\end{equation}
can be written, respectively, as
\begin{equation}
\label{chi_eta}
\bm \chi =
{1\over \sqrt{2}}
\left[
\begin{array}{c}
e^{-i\eta /2}  \\ 
\pm e^{i\eta /2} 
\end{array}
\right].
\end{equation}
Comparing this with the textbook formula of an SU(2) spinor,
\begin{equation}
\label{JJS}
|\hat{\bm n} \pm \rangle 
= \left[
\begin{array}{c}
e^{-i\eta /2} \cos (\theta /2) \\ 
\pm e^{i\eta /2} \sin (\theta /2) 
\end{array}
\right],
\end{equation}
pointed in the direction of a unit vector $\hat{\bm n}$ 
specified by a polar angle $\theta$ and an azimuthal angle $\eta$
of the SU(2) spin space: $(s_x, s_y, s_z)$,
one can verify
that the surface effective spin is {\it locked} in the $(s_x, s_y)$-plane,
i.e., $\theta = \pi /2$.

\begin{figure}
\begin{center}
\includegraphics[width=8cm]{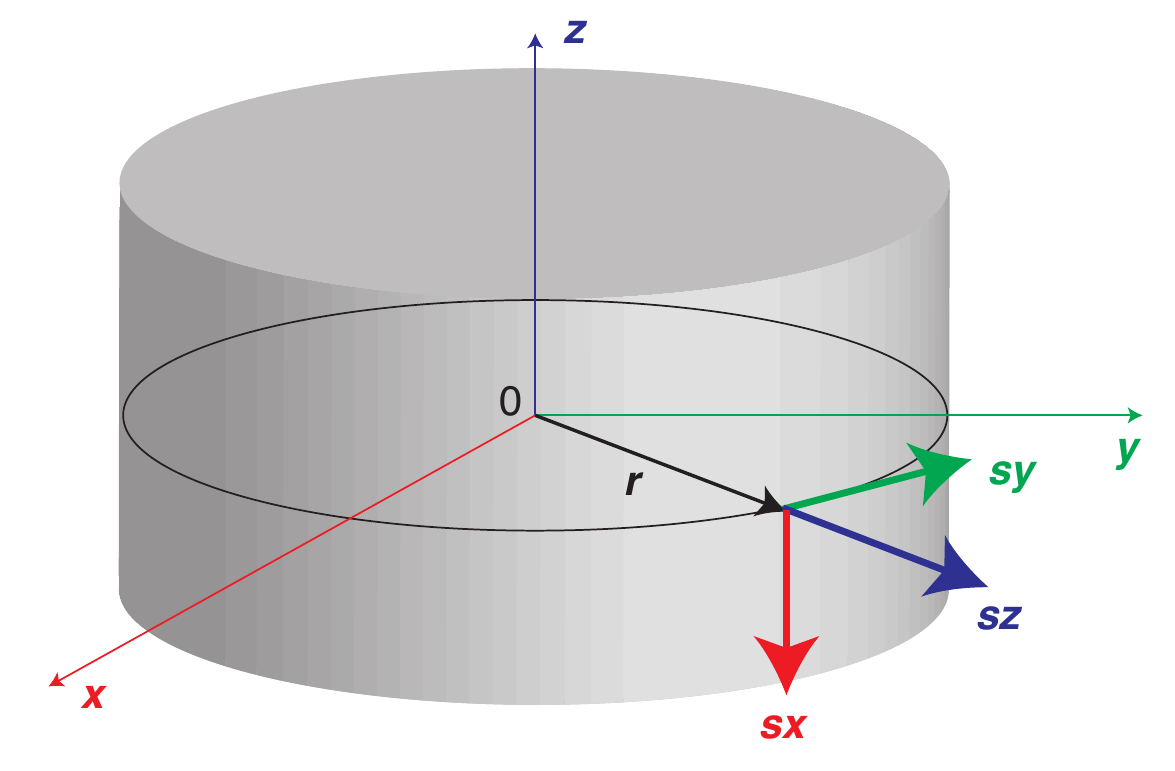}
\end{center}
\caption{Spin coordinates on the cylindrical surface.}
\label{spec_screw}
\end{figure}

In the derivation of Eq. (\ref{ham_surf}),
we have chosen
the spin quantization axis in the direction of $\hat{\bm r}$;
see Eq. (\ref{r_pm}).
Then, on which plane is the surface $(c_1, c_2)$-spin actually locked?
In accordance with Eq. (\ref{r_pm}),
the spin-space coordinates should be redefined as
\begin{eqnarray}
\hat{\bm s}_x:\ \ \hat{\bm x} &\rightarrow& -\hat{\bm z},
\nonumber \\
\hat{\bm s}_y:\ \ \hat{\bm y} &\rightarrow& \hat{\bm \phi},
\nonumber \\
\hat{\bm s}_z:\ \ 
\hat{\bm z} &\rightarrow& \hat{\bm r},
\end{eqnarray}
on the cylindrical surface (see FIG. 2),
since $k_\phi$ ($k_z$) play the role of $k_y$ ($- k_x$),
where $\hat{\bm \phi} = (- \sin \phi, \cos \phi)$.
Taking this into account, one can interpret Eq. (\ref{ham_surf_w/o}),
neglecting the anisotropy ($B=A$), as
\begin{equation}
H_{2D} ^{(\bm \chi)} \sim A (\bm \sigma \times \bm k)_z.
\end{equation}

In this regard, one can view Eq. (\ref{sol_bc_1}) 
with coefficients $\chi_1$ and $\chi_2$ given in Eq. (\ref{chi_eta})
as an SU(2) spin state,
\begin{equation}
\label{spin_eta}
|\bm r \pm\rangle = \chi_1 |\bm r +\rangle + \chi_2 |\bm r -\rangle,
\end{equation}
decorated by an accompanying orbital degree of freedom.
Eq. (\ref{spin_eta}) 
represents a spin state
which is locked in the plane perpendicular to $\hat{\bm r}$;
unit normal vector of the cylindrical surface.
This is a clear indication that the effective spin
on the surface of a topological insulator cylinder
is a real spin which is locked at each point of the cylinder
in-plane to its tangential surface [FIG. 1, panel (a)],
unlike the sub-lattice pseudo-spin on the cylindrical surface of a carbon nanotube.
\cite{R}

One step backward, notice that this peculiar property of the spin Berry phase $\pi$
manifest in Eqs. (\ref{ham_surf}) and (\ref{SGT})
is here not derived from the property of the effective 
Dirac Hamiltonian, i.e., Eq. (\ref{ham_surf_w/o}) on the cylindrical coordinates.
It was encoded in the dependence of eigenspinors (\ref{sol_bc})
on the spatial angle $\phi$, and the explicit form of $H_\parallel$
as given in Eq. (\ref{ham_para}).
In Refs. \cite{Vishwanath_PRB, Vishwanath_PRL,Mirlin}
the same conclusion was drawn
by observing the surface effective Hamiltonian on a curved surface.
Here, the Dirac equation was derived simultaneously with the spin Berry phase;
the existence of the spin Berry phase is indeed {\it encoded} in the bulk 
3D Hamiltonian.

\subsection{Cylindrical nanowire perpendicular to the crystal growth axis}

We have considered so far electronic states on the surface 
of a cylinder whose axis of symmetry is pointed along the $c$-axis,
i.e., in the direction of crystal anisotropy.
Since the rotational symmetry around the $c$-axis is presumed,
any tangential surface of the cylinder is equivalent.
Under such circumstances, we have seen explicitly that
the effective spin degree of freedom appearing in the Dirac equation
(\ref{ham_surf}) is constrained to the curved (cylindrical) surface.

Here, we consider a less trivial case of broken rotational symmetry, i.e.,
with the cylindrical axis chosen perpendicular to the $c$-axis.
Of course, nanowires are not cylindrical in real samples, but
have several surfaces.
It is also unlikely that the axis of the wire is
perfectly aligned with the axis of crystal symmetry ($c$-axis).
In the presence of such rotational anisotropy, it is less trivial
whether the effective spin degree of freedom on the surface is always
tied to the curved surface.

To implement an anisotropic cylindrical surface
we consider here the case of the crystal $c$-axis 
pointed in the $x$-direction,
keeping the symmetry axis of the cylinder pointed always in the
$z$-direction;
obviously, one can equally rotate the cylinder in the $y$-direction
with keeping the crystal growth axis in the $z$-direction.
The bulk effective Hamiltonian as Eq. (\ref{ham_bulk_symb})
for such a rotated crystal becomes (here, we do rotate the crystal),
\begin{equation}
\label{ham_bulk_2}
H_{3D} = M \tau_z + \tau_x (B \sigma_x k_x + A \sigma_y k_y + A \sigma_z k_z ).
\end{equation}

To identify the surface electronic states which span the
basis for the 2D surface Dirac Hamiltonian,
we introduce the same cylindrical coordinate
as Eq. (\ref{theta_def}) and decompose the Hamiltonian (\ref{ham_bulk_2})
into perpendicular ($H_\perp$) and parallel ($H_\parallel$) 
components in parallel with Eqs. (\ref{decomp}),
(\ref{ham_perp}) and (\ref{ham_para}).
Some parameters need, of course, redefinition or exchange.
Let us first focus on
\begin{equation}
\label{ham_perp_2}
H_\perp = M_\perp \tau_z + k_r \tau_x (B \sigma_x \cos \phi + A \sigma_y \sin \phi).
\end{equation}
To diagonalize the real spin part of the Hamiltonian (\ref{ham_perp_2}),
it is convenient to introduce an auxiliary angle $\tilde{\phi}$, defined as,
\begin{eqnarray}
\label{theta_tilde}
B \cos \phi = \tilde{A} \cos \tilde{\phi},
\nonumber \\
A \sin \phi =  \tilde{A} \sin \tilde{\phi},
\end{eqnarray}
where 
\begin{equation}
\label{A_tilde}
\tilde{A}=\tilde{A} (\phi)=\sqrt{B^2 \cos^2 \phi +A^2 \sin^2 \phi}.
\end{equation}
In analogy with Eq. (\ref{ham_perp_pm}),
one can partially diagonalize Eq. (\ref{ham_perp_2}) as
\begin{equation}
\label{ham_perp_pm_2}
H_\perp |\tilde{\bm r}  \pm \rangle = 
\left( M_\perp \tau_z \pm \tilde{A} k_r \tau_x \right) |\tilde{\bm r} \pm \rangle,
\end{equation}
but here the eigenspinors represent no longer spin states in the $\hat{\bm r}$-direction
normal to the surface of the cylinder.
The new eigenspinors $|\tilde{\bm r} \pm \rangle$
are formally analogous to $|\bm r \pm \rangle$
defined as in Eqs. (\ref{r_pm}), but pointed in the
direction specified by $\tilde{\phi}$
introduced above;
\begin{equation}
\label{r_tilde}
|\tilde{\bm r} + \rangle = {1\over \sqrt{2}}
\left[
\begin{array}{c}
1 \\ e^{i\tilde{\phi}} 
\end{array}
\right],\ \ 
|\tilde{\bm r} - \rangle =
 {1\over \sqrt{2}}
\left[
\begin{array}{c}
1 \\ - e^{i\tilde{\phi}} 
\end{array}
\right].
\end{equation}
The remaining procedure is perfectly in parallel with the previous case.
After imposing the boundary condition on the cylindrical surface of the
topological insulator, one finds as the basis spinors
for constructing the surface effective Hamiltonian,
\begin{eqnarray}
|\psi \rangle &=&
\left[
c_1 |\lambda^{(+)} +\rangle |\tilde{\bm r} +\rangle +
c_2 |\lambda^{(+)} -\rangle |\tilde{\bm r} -\rangle
\right]
\rho(r)
\label{sol_bc_symb_2}
\\
&=&\left[
{c_1 \over 2}
\left[
\begin{array}{c}
1 \\ i \\ e^{i\tilde{\phi}} \\ i e^{i\tilde{\phi}}
\end{array}
\right]
+
{c_2 \over 2}
\left[
\begin{array}{c}
1 \\ -i \\ - e^{i\tilde{\phi}} \\ i e^{i\tilde{\phi}}
\end{array}
\right]
\right]
\rho (r)
\nonumber \\
&\equiv&
c_1 |\tilde{\bm r} +\rangle\rangle + c_2 |\tilde{\bm r} - \rangle\rangle,
\label{sol_bc_an}
\end{eqnarray}
where $\rho (r)$ is given as in Eq. (\ref{rho_r}) 
with a normalization factor expressed in terms of
$\lambda_\pm$ given as in Eq. (\ref{lambda_pm}) 
but with $A$ replaced by $\tilde{A}$ given in Eq. (\ref{A_tilde}). 


To find the surface effective Hamiltonian, we calculate again
the matrix elements of $H_\parallel$,
here, in terms of $|\tilde{\bm r} \pm \rangle\rangle$
associated with
$|\tilde{\bm r} \pm \rangle$
introduced in Eqs. (\ref{r_tilde}).
We have introduced the parameter $\tilde{\phi}$ in Eqs. (\ref{theta_tilde})
and was able to construct the basis spinors
$|\tilde{\bm r} \pm \rangle\rangle$
which successfully span the subspace representing the
solution of the boundary problem.
$\tilde{\phi}$-parametrization was useful, since $H_\perp$ shows a nice
transformation property in terms of $\tilde{\phi}$.
It is, however, no longer the case for $H_\parallel$, which reads explicitly
in the present case,
\begin{eqnarray}
H_\parallel &=& M_\parallel \tau_z + A \tau_x \sigma_z k_z 
\nonumber \\
&+& \tau_x (- B \sigma_x \sin \phi + A \sigma_y \cos \phi) k_\phi
\\
&=&
\left[
\begin{array}{cccc}
M_\parallel & A k_z & 0 & -i c_\phi k_\phi \\
A k_z & -M_\parallel  & -i c_\phi k_\phi & 0 \\
0 & i c_\phi^* k_\phi & M_\parallel & -A k_z \\
i c_\phi^* k_\phi & 0 & -A k_z & -M_\parallel
\end{array}
\right],
\end{eqnarray}
where we have introduced
\begin{equation}
c_\phi= A \cos \phi - i B \sin \phi.
\end{equation}
$c_\phi$ is a complex number
which is a function of the real parameter $\phi$.
Note that $c_\phi$ does not simplify with the use of $\tilde{\phi}$.

To derive the surface effective Hamiltonian $H_{2D}$,
one calculates again the matrix elements,
$\langle\langle \tilde{\bm r} \pm | H_\parallel |\tilde{\bm r}\pm \rangle\rangle$
in terms of the new $\tilde{\bm r}$-basis.
One can explicitly verify that diagonal elements of the surface effective Hamiltonian
vanish, i.e.,
\begin{equation}
\tilde{H}_{2D} =
\left[
\begin{array}{cc}
0 & \langle\langle \tilde{\bm r} +| H_\parallel |\tilde{\bm r}- \rangle\rangle
\\ 
\langle\langle \tilde{\bm r} -| H_\parallel |\tilde{\bm r} + \rangle\rangle & 0
\end{array}
\label{ham_surf_2}
\right].
\end{equation}
The off-diagonal elements involve a first-order derivative with respect to $\phi$,
yielding a Berry phase term; e.g.,
\begin{eqnarray}
&&\langle\langle \tilde{\bm r} +| H_\parallel |\tilde{\bm r}- \rangle\rangle =
\nonumber \\
&&
-i A k_z
+{1\over 2}\left(c_\phi k_\phi e^{i\tilde{\phi}} + e^{-i\tilde{\phi}} c_\phi^* k_\phi\right).
\label{berry_aniso}
\end{eqnarray}
Note that 
terms coming from the $\phi$-dependence of $\lambda_\pm$,
given as in Eq. (\ref{lambda_pm}),
cancel when integrated over $r$.
The $\phi$-dependence exists not only in the explicit $r$-dependence 
of $\rho (r)$ on $\lambda_\pm$; cf. Eq. (\ref{rho_r}),
but also in the normalization factor of $\rho (r)$. 
The $\phi$-dependence of $\lambda_\pm$ stems from that of $\tilde{A}=\tilde{A} (\phi)$. 
In Eq. (\ref{berry_aniso})
the Berry phase term appears in the second part 
when $k_\phi$ acts on the exponent of $e^{i\tilde{\phi}}$, 
and is found to be of the form,
\begin{equation}
\label{Omega}
\Omega (\phi)=
{c_\phi e^{i\tilde{\phi}} \over 2R} {d \tilde{\phi} \over d\phi}
= {c_\phi e^{i\tilde{\phi}} \over 2R} {AB \over \tilde{A}^2}.
\end{equation}
Notice that only the real part of this factor influences the phase 
of the wave function and is identified as the local spin Berry phase.
This phase factor can be eliminated from the eigenvalue equation:
$\tilde{H}_{2D} \bm c = E_\parallel \bm c$,
by employing a singular gauge transformation 
analogous to Eq. (\ref{SGT}), i.e., by
$\bm c = e^{-i\tilde{\phi}/2} \bm \chi$.
Since $\phi$ and $\tilde{\phi}$ have the same winding property, 
\begin{equation}
\tilde{\phi}(\phi+2\pi)-\tilde{\phi}(\phi)=2\pi,
\end{equation}
the eigenspinor, Eq. (\ref{chi})
is precisely {\it double-valued} with respect to a $2\pi$-rotation
of $\phi$, i.e., the spin Berry phase is just $\pi$.

The off-diagonal elements of Eq. (\ref{ham_surf_2}) are 
also susceptible of the imaginary part of $\Omega (\phi)$ given as in Eq. (\ref{Omega}).
This imaginary part introduces a non-uniform modulation into 
the amplitude of the wave function of surface electronic states.
Such a situation is somewhat similar to the case of a WKB wave function describing
the system with a spatial non-uniformity.
The phase of such WKB wave function is expressed in terms of
a ``wave vector'' $k$ which depends on the space coordinate.
Meanwhile its amplitude shows also a space dependence.
In the present formulation, this appears as an imaginary part of
$\Omega (\phi)$.

Let us come back to the question,
``In which direction is the eigenspinor
$\bm c$, or equivalently $\bm \chi$,
pointed in the spin space?''
Taking it into account that
our surface effective Hamiltonian is defined on a space spanned
by $ |\tilde{\bm r}\pm \rangle\rangle$, 
we redefine the spin coordinates as,
\begin{eqnarray}
\hat{\bm s}_x:\ \ \hat{\bm x} &\rightarrow& -\hat{\bm z},
\nonumber \\
\hat{\bm s}_y:\ \ \hat{\bm y} &\rightarrow& \tilde{\bm \phi},
\nonumber \\
\hat{\bm s}_z:\ \ \hat{\bm z} &\rightarrow& \tilde{\bm r},
\end{eqnarray}
where $\tilde{\bm \phi} = ( -\sin\tilde{\phi}, \cos\tilde{\phi})$.
This means that the eigenstate $\bm c$ of $\tilde{H}_{2D}$ given in Eq. (\ref{ham_surf_2})
can be viewed as an SU(2) spin state,
\begin{equation}
\label{spin_tilde}
|\tilde{\bm r} \pm\rangle = b_1 |\tilde{\bm r} +\rangle + b_2 |\tilde{\bm r} -\rangle,
\end{equation}
accompanied by orbital spins.
Eq. (\ref{spin_tilde}) represents a spin state locked in the plane
perpendicular to $\tilde{\bm r}$.
Note that $\tilde{\bm r}$ is {\it not} a unit vector {\it normal} to the cylinder
surface.
Thus in the case of broken cylindrical symmetry,
the surface effective spin can have a finite component normal to 
a tangential surface of the cylinder [FIG. 1, panel (b)].
An out-of-plane component of the surface effective spin
appears also as a consequence of the hexagonal warping,
\cite{Fu, Xu, Ando}
and in a system of topological insulator quantum dot. 
\cite{Thierry}

In this section, 
we have examined the electronic states on 
cylindrical surfaces of an anisotropic topological insulator.
An explicit one-to-one correspondence between 
the effective spin in the surface Dirac Hamiltonian
and the real spin inherent to the 3D bulk effective Hamiltonian
has been established.
The existence of spin Berry phase $\pi$ has been unambiguously
shown in this context.
In the cylindrically symmetric case (Sec. II A), i.e.,
in the case of a cylindrical nanowire parallel to the crystal $c$-axis,
we have shown explicitly that
the effective surface spin is constrained in-plane to 
the real-space tangential plane of the cylinder.
In the absence of such cylindrical symmetry
(Sec. II B: case of a cylindrical nanowire perpendicular to the crystal $c$-axis)
the effective surface spin can have a finite amplitude in the direction
normal to the cylindrical surface.
However, when the reference point on the cylindrical surface travels
once around the cylinder (i.e., winds the cylinder once), 
the resulting spin Berry phase is indeed $\pi$.
This point will be further confirmed in the numerical experiments in Sec. III.

\begin{figure}
\begin{center}
\includegraphics[width=8cm]{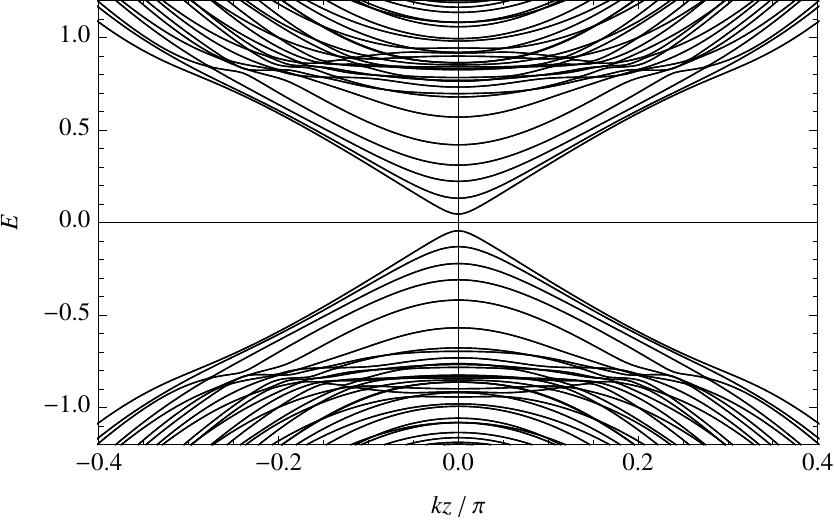}
\includegraphics[width=8cm]{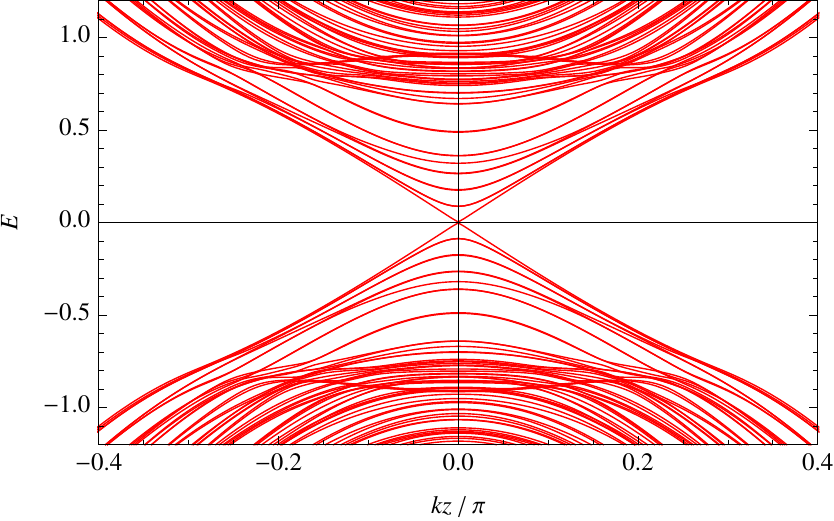}
\end{center}
\caption{Energy spectrum $E(k_z)$ of a topological insulator nanowire 
with a rectangular cross section ($N_x=12$, $N_y =16$).
The horizontal axis $k_z$ represents
the crystal momentum along the wire.
The crystal anisotropy is introduced in the direction
perpendicular to the wire (in the $x$-direction, $B/A=0.7$).
$M_0/M_2= -1$
(STI with a surface Dirac cone at the $\Gamma$-point).
The lower panel is for 
the same nanowire pierced by a $\pi$-flux tube.
To avoid formation of a bound state,
the $\pi$-flux is divided into two $\pi /2$-flux tubes
penetrating two neighboring plaquettes in the wire center region.
}
\label{spec_pi}
\end{figure}

\section{Energy spectrum in the presence of a $\pi$-flux tube}

The spin Berry phase appearing in the surface Dirac Hamiltonian;
cf. Eq. (\ref{ham_surf}),
is often discussed
\cite{Vishwanath_PRB, Vishwanath_PRL, Mirlin, Moore, Franz, ITT} 
in the context of Aharonov-Bohm experiment 
on topological insulator nanowires.
\cite{AB_exp}
Let us first recall that the existence of a spin Berry phase $\pi$
leads, on the surface of a cylindrical nanowire,
to the appearance of a finite-size energy gap 
in the spectrum of surface electronic states.
The spin Berry phase $\pi$ modifies the periodic boundary condition
around the wire to an anti-periodic boundary condition,
leading to opening of the gap.
Let us explicitly see this.
The wave function of such a surface electronic state is 
an eigenstate of Eq. (\ref{ham_surf}), which is a plane wave,
\begin{equation}
\psi (z, \phi) = e^{i k_z z} e^{i k_\phi R\phi}.
\label{pw}
\end{equation}
The corresponding eigenenergy reads
\begin{equation}
E(k_z, k_\phi) = \pm A \sqrt{k_z^2 + k_\phi^2}.
\end{equation}
Here, we neglect the anisotropy ($B=A$).
The energy spectrum $E(k_z)$ of the surface electronic states
is determined by imposing
(anti-)periodic boundary condition to Eq. (\ref{pw}).
The spin Berry phase $\pi$ replaces the periodic boundary condition:
\begin{equation}
\psi (z, \phi + 2\pi) = \psi (z, \phi),
\end{equation}
by an anti-periodic boundary condition,
\begin{equation}
\psi (z, \phi + 2\pi) = - \psi (z, \phi),
\end{equation}
shifting the allowed values of $k_\phi R$ from integers 
($k_\phi R=0, \pm 1, \pm 2, \cdots$)
to half odd integers ($k_\phi R=\pm 1/2, \pm 3/2, \cdots$).
Importantly, the $k_\phi =0$ and correspondingly the zero-energy bound
state was purged from the lowest energy portion of the spectrum
by this $\pi$-phase shifting
(cf. FIG. 3, upper panel).
In the Aharonov-Bohm geometry, this finite-size energy gap
associated with the spin Berry phase
is compensated by an Aharonov-Bohm phase $\phi_{AB}$, and in some cases
we expect the energy spectrum closes its gap.

In the previous section the spin Berry phase $\pi$ for surface
electrons is derived in the presence and absence (due
to crystal anisotropy) of cylindrical symmetry, that is, regardless of
fulfilling the spin-to-surface locking condition. 
Here, we confirm numerically the robustness of
the $\pi$ spin Berry phase against cylindrical symmetry breaking,
using a tight-binding model for a topological insulator nanowire.
We attempt to show 
that the amount of the "integrated" Berry phase,
corresponding to a $2\pi$-rotation of the azimuthal angle $\phi$,
is precisely $\pi$.
To verify this explicitly, we introduce a $\pi$ magnetic flux tube
piercing the nanowire, as a {\it probe},
and investigate the corresponding energy spectrum.
Exact cancellation of the spin Berry phase and the Aharonov-Bohm phase
is confirmed by observing the closing of finite-size energy gap.

To implement a "shape" in real space such as a nanowire
we consider in the following a tight-binding version of
Eq. (\ref{ham_bulk_2}) on a cubic lattice.
We first make the following replacement:
\begin{equation}
k_j \rightarrow \sin k_j,
\end{equation}
where $j=x,y,z$, for the $k_j$'s in Eq. (\ref{ham_bulk_2}),
and similarly,
\begin{equation}
k_j^2 \rightarrow 2(1-\cos k_j),
\end{equation}
for $k_j^2$'s in $M$.
After this replacement the Hamiltonian  (\ref{ham_bulk_2})
can be interpreted as a tight-binding Hamiltonian, 
i.e.,
\begin{eqnarray}
\label{ham_tb}
H_{3D} &=& \sum_{x, y, z} \big\{
(M_0 + 6 M_2) |x, y, z\rangle \langle x, y, z| 
\nonumber \\
&+& \big(
t_x |x+1, y, z\rangle \langle x, y, z| + t_y |x, y+1, z\rangle \langle x, y, z| 
\nonumber \\
&+& t_z |x, y, z+1\rangle \langle x, y, z| + h.c.
\big)
\big\},
\label{ham}
\end{eqnarray}
where
\begin{eqnarray}
t_x &=& i {B \over 2} \tau_x \sigma_x - M_2 \tau_z,\ \ 
t_y = i {A \over 2} \tau_x \sigma_y - M_2 \tau_z.
\nonumber \\
t_z &=& i {A \over 2} \tau_x \sigma_z - M_2 \tau_z.
\end{eqnarray}
Depending on the value of $M_0/M_2$,
the tight-binding Hamiltonian (\ref{ham_tb}) describes either
strong/weak topological insulators (STI/WTI) or an ordinary insulator;
$-4<M_0/M_2<0$ and $-12<M_0/M_2<-8$ $\rightarrow$ STI,
$-8<M_0/M_2<-4$ $\rightarrow$ WTI, 
$M_0/M_2<-12$ and $0<M_0/M_2$ $\rightarrow$ ordinary insulator.
In the following, we will mainly focus on the case
$-4<M_0/M_2<0$, corresponding to STI with a surface Dirac cone at the $\Gamma$-point.
So far the Hamiltonian is translationally symmetric.
We now restrict the electrons to move only inside the nanowire
with a rectangular cross section:
$1 \le x \le N_x$, $1 \le y \le N_y$.
This can be done simply by switching off unnecessary hopping amplitudes.
After this only $k_z$ remains to be a good quantum number;
we always assume a periodic boundary condition in the $z$-direction.

\begin{figure}
\begin{center}
\includegraphics[width=8cm]{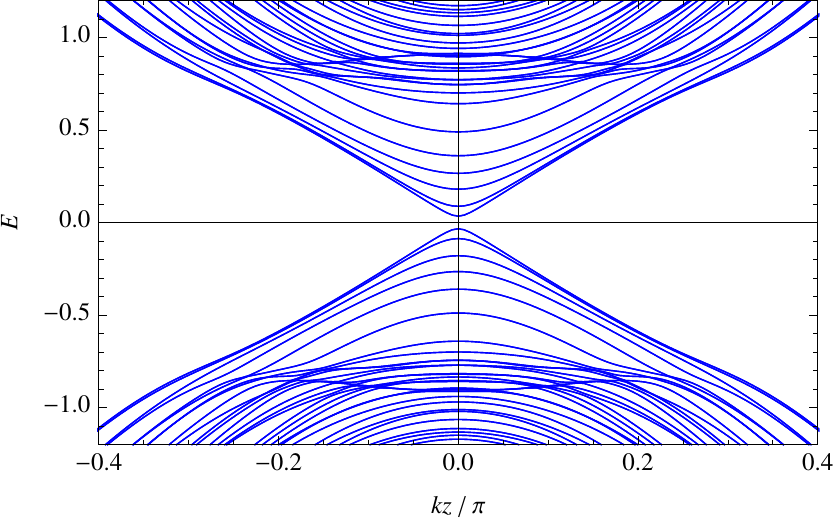}
\includegraphics[width=8cm]{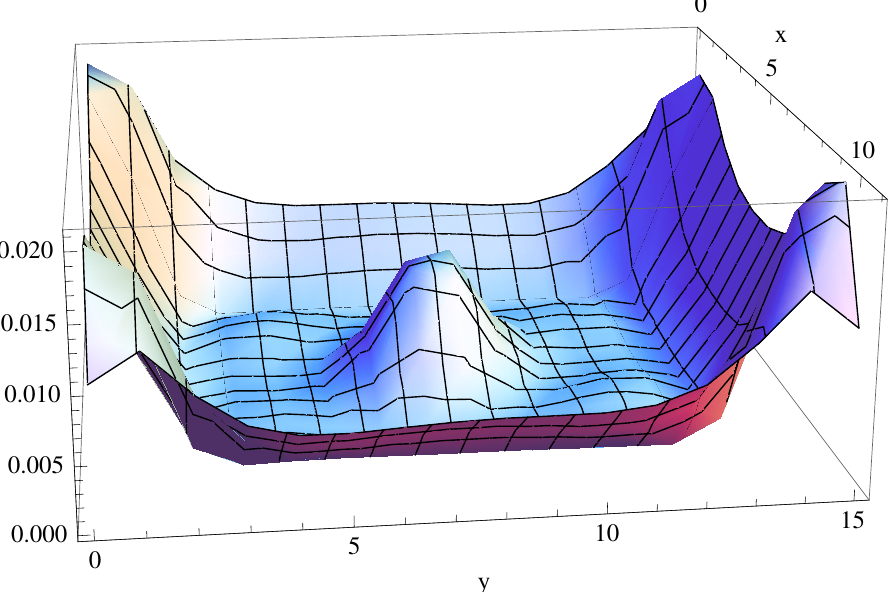}
\end{center}
\caption{Energy spectrum $E(k_z)$ and a typical shape of
the (squared amplitude of) the lowest energy wave function
in the presence of a $\pi$-flux tube
penetrating a series of single plaquettes along
the axis of nanowire.
The cross section is rectangular, $N_x=12$, $N_y = 16$,
the anisotropy is $B/A=0.7$.
$M_0/M_2= -1$. $M_2=A=1$.
The wave function is depicted at $k_z/\pi =0.03$.
}
\label{wf_pi}
\end{figure}

We then introduce  an Aharonov-Bohm flux $\phi_{AB}$
piercing the nanowire.
We consider typically the case of $\phi_{AB}=\pi$
(case of a $\pi$-flux tube).
The simplest way to do this
is to let a $\pi$-flux tube penetrate a central plaquette of
each $z$-layer, e.g., the plaquette centered at
$(x,y)=(N_x /2+1/2, N_y /2+1/2)$ for $N_x$ and $N_y$ being an even integer.
The insertion of a flux can be achieved by Peierls substitution,
e.g., $t_x \rightarrow t_x e^{i \phi_{AB}}$
for $x=N_x /2 \rightarrow N_x /2 +1$,
$y=1,\cdots, N_y/2$.
This turns out, however, to be not the best solution for us, since
a $\pi$-flux tube penetrating a {\it single} plaquette involves
a (zero-energy) bound state.
\cite{Franz, Zaanen}
In the nanowire geometry
such a bound state appears as the lowest energy
helical modes bound to the flux tube,
propagating in the direction opposite to the preformed
surface electronic states,
i.e., electrons localized on the surface of the cylinder.
Formation of such a pair of {\it counter-propagating} 
modes separated only by a finite distance leads
naturally to gap opening (FIG. 4, upper panel);
mixing of the two counter-propagating modes
(see the wave function, in FIG. 4, lower panel)
causes level repulsion between the two initially gapless
states.

To avoid formation of such a bound state 
we rather introduce here
a total magnetic flux $\pi$ divided into two $\pi /2$-flux tubes 
penetrating the two neighboring plaquettes, e.g.,
plaquettes centered at 
$(x,y)=(N_x /2 -1/2, N_y /2 +1/2)$ and $(N_x /2 +1/2, N_y /2 +1/2)$.
With this we could see a clear signature of the finite-size gap closing
(FIG. 3, lower panel).
There appears no low-energy bound states around a
$\pi /2$-flux tube.
In the same figure one can actually recognize such bound states pushed
up into the high-energy spectrum.

In the above example,
by changing the amount of Aharonov-Bohm flux
$\phi_{AB} = \pi, 2\pi, 3\pi, \cdots$,
we find an even/odd feature 
(gapped, gapless, gapped, gapless, $\cdots$)
in the energy spectrum of surface electronic states.
An alternative way
to verify explicitly that the amount of this Berry phase is precisely $\pi$
is to investigate a similar even/odd feature 
due to crystal dislocation lines
(results not shown here).
We have shown previously that the electronic states along
such a dislocation or equivalently a nanowire
exhibits a finite size energy gap,
manifesting the existence of spin Berry phase.
\cite{ITT}

What have we verified in these numerical simulations?
(Especially in its relation to what we have discussed in Sec. II B.)
In the example we have presented in this section
the cylinder itself is distorted,
having a rectangular cross section.
In addition to such a structural asymmetry,
we have also taken into account the crystal anisotropy.
Our data indicate that the global spin Berry phase $\pi$
is robust under the coexistence of a structural asymmetry
and a crystal anisotropy.

\section{Conclusions}
We have studied the electronic states on cylindrical surfaces of 
an anisotropic topological insulator.
We have established an explicit one-to-one correspondence between 
the effective spin in the surface Dirac Hamiltonian
and the real spin inherent to the 3D bulk effective Hamiltonian.
The effective spin on the surface of a topological insulator has a 
property of being constrained on its tangential surface, and
in particular, when its surface is warped (e.g., into a cylinder)
the effective spin feels this change of tangential plane in real space,
and consequently,
the effective spin completes a $2\pi$ rotation when the reference point
travels once around the cylinder.
The existence of spin Berry phase $\pi$ is naturally understood in this context.
However, 
on the surface of a topological insulator with broken cylindrical symmetry 
in the presence of crystal anisotropy,
the effective spin does not follow {\it locally} the tangential plane in real space,
i.e., the effective spin can have generally a component normal to the surface.
This has been shown analytically in Sec. II B, using a rather simple model,
Whereas, {\it globally} the spin Berry phase $\pi$ is robust against anisotropy
and breaking of the cylindrical symmetry.
The latter has been verified numerically in Sec. III.

In the examples we have considered in this paper,
the ``curved'' surfaces did not really have a curvature;
a cylindrical surface is flat in the proper use of terminology in differential geometry
(its Riemann curvature is null).
On a genuinely curved surface with a finite Riemann curvature, e.g., on a sphere,
the concept of $\pi$ spin Berry phase may need some modification or a generalization.
Naively, it is expected to involve a solid angle associated with parallel transport on a sphere.
We leave future studies a more rigorous discussion on such an issue.

\acknowledgments
The authors are supported by KAKENHI;
KI and AT under ``Topological Quantum Phenomena''
[Nos. KI: 23103511, AT: 23103516],
YT and AT by a Grant-in-Aid for Scientific Research (C) 
[Nos. YT: 21540389, AT: 23540461].

\bibliography{k2_r2_v6}

\end{document}